\documentclass[prl,twocolumn]{revtex4}
\usepackage{amsmath}
\usepackage{amssymb}
\usepackage{amsfonts}
\usepackage[]{graphicx}

\begin{document}

\title{Measuring the quantum states of a mesoscopic SQUID using a small Josephson junction}

\author{Ren\'{e} Lindell, Jari Penttil\"{a}, Mika Sillanp\"a\"a, and Pertti Hakonen}
\affiliation{Low Temperature Laboratory, \\
Helsinki University of Technology,\\
FIN-02015 HUT, Finland \\
}
\date{\today}

\begin{abstract}

We have experimentally studied the energy levels of a mesoscopic SQUID using inelastic Cooper-pair tunneling.
The tunneling in a small Josephson junction depends strongly on its electromagnetic environment.
We use this fact to do energy level spectroscopy of a SQUID-loop by coupling it to a small junction.
Our samples with strong quasiparticle dissipation are well described by a model of a particle localized in one of the
dips in a cosine-potential, while in the samples with weak dissipation we can see formation of energy bands.
\end{abstract}
\maketitle


\newpage

The Josephson junction, despite its simple structure has proven to be surprisingly versatile and
new applications are found in quantum computing and nanoelectronics \cite{vion,delahaye}.
The devices are based on the quantum mechanical behavior of the superconducting phase variable \cite{Clarke},
which has been previously studied with either rf-irradiation \cite{Devoret85} or during
rapid current ramping \cite{Silvestrini97, wallraff}. We are using a different probe, namely, an additional
mesoscopic Josephson junction.  Our scheme is based on the theory of phase fluctuations \cite{Devoret90,IN}, according to which Coulomb
blockade in a single superconducting tunnel junction is strongly affected by its environment. Non-coherent
Cooper pair tunneling is allowed only if energy is exchanged with the surroundings. Thus, this inelastic Cooper
pair tunneling provides a good tool for observing all kinds of environmental modes in a rather simple fashion  \cite{Holst}.

In this Letter we present detailed spectroscopic investigations on small SQUID loops, which are driven from the nearly classical
limit ($E_J/E_C \gg 1$) deep into the quantum regime ($E_J/E_C \sim 1$). Our results yield evidence for higher energy bands
of the macroscopic phase variable in a regime ($E_J/E_C \gtrsim 1$) where they have not been investigated before \cite{Flees}. In addition, our
experiment provides the first verification that multiphoton transitions between the levels of a quantum mechanical harmonic oscillator play a
prominent role in electron tunneling in a mesoscopic tunnel junction.

As an energy detector in our measurement we use a voltage biased, superconducting tunnel junction
which has a smaller size and critical current than the junction we want to study.
For large (conventional) Josephson junctions the supercurrent is given by $I=I_c \sin(\varphi)$, where $I_c$ is the critical current, which
is related to the Josephson coupling energy $E_J = \hslash I_C/(2e)$. The phase $\varphi(t) =\int_{-\infty}^{t} \frac{2e}{\hslash }V(t')dt'$
is defined as an integral of the voltage $V$ across the tunnel barrier. For small junctions, where the charging energy
$E_C = e^2/(2C) \gg E_J$ Cooper pair tunneling is inelastic and given by
\begin{equation}
{I(V)}={\frac{\pi eE_{J}^{2}}{\hslash }}\left[ P(2eV)-P(-2eV)\right],
\label{IV}
\end{equation}
where $P(E)$ is a function describing the probability of energy exchange between a tunnel junction
and its electromagnetic environment and depends on the impedance seen by the junction \cite{IN}.

At low temperatures, the junction environment, i.e. the heat bath,
is in its ground state and $P(E) \simeq 0$ for $E<0$. Thus, the latter term in Eq. (\ref{IV})
can be neglected and ${I(V)}$ becomes directly proportional to $P(2eV)$. The theory is valid
for linear impedances constructed from lumped elements. Nevertheless, we argue that the idea of
energy exchange can be generalized so that a discrete spectrum of energy levels in the
environment will cause a set of discrete peaks in the $IV$-curve. Hence, the small detector junction
can be used for spectroscopy.

A Josephson junction can be described by the Schr\"odinger equation \cite{LZ}
\begin{equation}
\frac{d^{2}\psi(\varphi) }{d(\varphi/2)^{2}}+\left( \frac{E}{E_{c}}+\frac{E_{J}}{E_{c}}%
\cos \varphi + \frac{I}{I_C} \varphi\right)\psi(\varphi) = 0 ,  \label{SCH}
\end{equation}
where $I$ is the current flowing through the junction. The current in our measurements always satisfies $I\ll I_C$, so
the tilt in the potential is negligible and setting $I=0$ in Eq.~\ref{SCH} leads to the familiar Mathieu-equation.
The single junction Hamiltonian can also be used to describe a SQUID-loop, where the loop size is
so small that the geometric inductance can be neglected and the loop is perfectly symmetric. The only difference
is that $E_J$ then depends periodically on an externally applied magnetic flux $\Phi$ according to
$E_J = 2E_J^{single} |\cos(\pi \Phi/\Phi_0)|^2$, where $\Phi_0 = h/(2e)$ and $E_J^{single}$ is the Josephson coupling
energy for a single junction.
For large $E_J/E_C$, the particle is trapped in one of the potential wells.
In this case, for currents $I \ll I_C$, the Josephson junction can be described by an inductance $L=\Phi _{0}/(2\pi I_{c})$.
Combined with the capacitance of the tunnel junction, the junction forms an
LC-oscillator with a characteristic resonance frequency of $\omega_{p}=1/\sqrt{LC}=\sqrt{8E_{J}E_{c}}/\hbar$.
Consequently, a Josephson junction behaves like a harmonic oscillator with a level spacing of
$\omega _{p}$. When $E_J/E_C$ becomes smaller, the energy levels are not harmonic but they will depend on the shape of the cosine-potential.

Depending on the environmental resistance seen by the Josephson junction, i.e. in our case
``the environment of the environment'', the junction can become completely delocalized and the whole periodicity
of the cosine-potential has to be accounted for \cite{Schon,LZ}. The eigenstates are then given by Bloch-functions
$\Psi_n(\varphi) = u_n(\varphi)e^{i\varphi q/(2e)}$, where $q$ is the quasi-charge, $n$ the band index and $u_n(\varphi)$ is a 2$\pi$-periodic
function. This phase transition from the localized to delocalized state happens when $R>R_Q$, where $R_Q =
h/(4e^2)$, or 6.45 k$\Omega$  \cite{Penttila, Yagi}. In our measurement we need a clear voltage-bias and thus we have not fabricated any resistor
close to the junction. The source of dissipation is, therefore, given by the quasiparticle resistance of the probe junction. This changes the
periodicity of the wave-functions from 2$\pi$ to 4$\pi$ and each band is split into two \cite{Schon}.
\begin{figure}[b]
\includegraphics[width=0.8\linewidth]{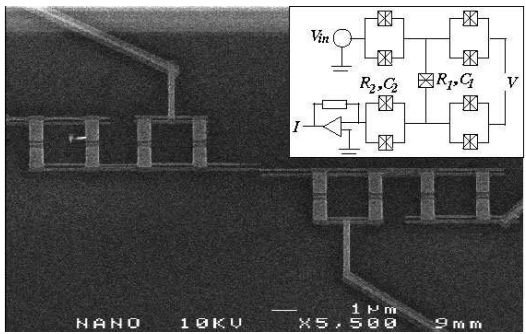}
\caption{\label{sem}SEM micrograph of a sample with 4 SQUIDs. The probe junction has an area of $100\times100$ $\rm{nm}^2$ and the
SQUID junctions $150\times550$ $\rm{nm}^2$ In the samples covered in this paper, additional gates leads were
available for the islands. Inset shows the schematic of the circuit in the 4-lead measurement.}
\end{figure}

We have carried out experiments with different circuit configurations; both 2- and 4-lead measurements including 1, 2 or
4 SQUID(s) coupled to a small detector junction. We will here describe measurements of two different samples; a 4-SQUIDs sample,
with four leads and a 1-SQUID sample with just two leads. A scanning electron micrograph of the 4-SQUID sample, together with a
schematic drawing of it, is shown in Fig. \ref{sem}. The SQUID configuration allows us to change the energy levels of the measured
system and enables us to resolve the resonances due to the SQUID(s) from other resonances in the environment. The critical current,
or equally, the value of $E_{J}$ could be tuned to less than 1\% of the maximum, which shows that our SQUIDs were very
homogeneous. The samples were made from aluminum with e-beam lithography and 2-angle evaporation in an UHV chamber.

\begin{table}[b]
\caption{Parameters for the 4-squid and 1-squid samples. Energies are given in units of $\mu$eV. The Ambegaokar-Baratoff
values for $E_J$ are given in parentheses.}
\label{table1}
\begin{ruledtabular}
\begin{tabular}{cccccc}
Sample&$R_T$ (k$\Omega$)&$C$(fF)&$E_J$&$E_C$&$E_J/E_C$\\
\colrule
4-SQUID (detector) & 166 & 0.5  & 3.6 & 160 & 0.023\\
4-SQUID (squid) & 2.5 & 7.6 & 544 (272) & 10.5 & 51.8\\
1-SQUID (detector)& 70 & 0.8 & 8.5 & 100 & 0.08\\
1-SQUID (squid)& 3.5 & 5.7 & 422 (188) & 14 & 30.1\\
\end{tabular}
\end{ruledtabular}
\end{table}

The four wire setup facilitates the determination of circuit parameters. The important parameters are $E_J$ and
$E_C$, or rather their ratio. The Ambegaokar-Baratoff (A-B) formula, $E_J = \hslash \pi \Delta /(4e^2 R_T)$, was used to find
$E_{J}$ from the normal state resistances, $R_T$,  while the capacitances where estimated from the junction areas (see Fig. 1) using a
value of 45 fF/$\mu m^2$ \cite{Haviland2}. The BCS-gap, $\Delta$, was about 215 $\mu$eV in our samples.
The experimental parameters for the different circuits are summarized in Table 1.

The samples were mounted into an rf-tight copper enclosure and cooled down to 80 mK with a plastic dilution refrigerator. The measurement
leads were filtered using 0.7 m long sections of Thermocoax. Mini-Circuits rf-filters with a cut-off frequency of 1.9 MHz
were employed on the top of the cryostat, at room temperature.

\begin{figure}[htc]
\includegraphics[width=0.8\linewidth]{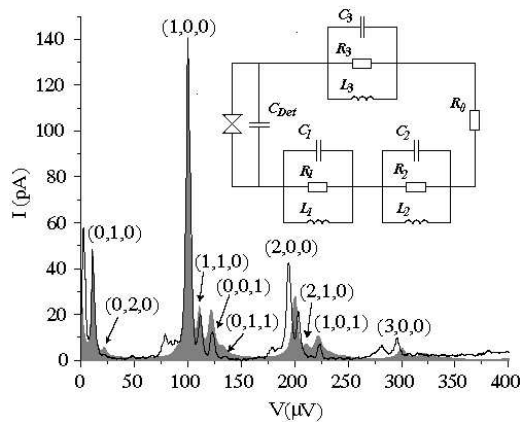}
\caption{\label{sjlc8iv}IV-curve for the 4-SQUID sample at maximum $E_J$ and the circuit model used in simulation. The full line shows the experimental
curve while the shaded area shows the simulated curve. The different excitations are denoted as (n,k,l), where n,k and l are the number of quanta excited.
The first index gives the resonance due to SQUIDs and the two other indices are due to other resonances in the circuit.
The parameters used in the simulation are $C_{Det}$ = 0.5 fF, $C_1$= 4 fF, $L_1$ = 2.28 nH, $R_1$ = 50 k$\Omega$ (SQUID),
$C_2$ = 0.5 pF, $L_2$ = 3.2 nH, $R_2$ = 30 k$\Omega$, $C_3$ = 2 fF, $L_3$ = 10.8 nH, $R_3$ = 3 k$\Omega$, $R_0$ = 100 $\Omega$ and $T$= 100 mK.}
\end{figure}

\begin{figure}[ht]
\includegraphics[width=0.8\linewidth]{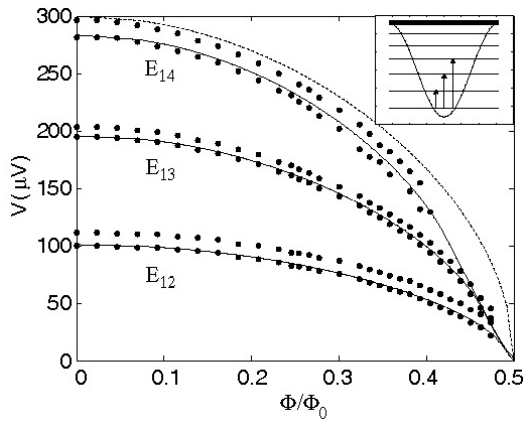}
\caption{\label{sjlc8p}The position of the main resonances as a function of applied flux for the 4-SQUID sample and the calculated transitions.
The main resonances due to the SQUID-loops consist of a double-peak structure. The potential well and the energy levels and shown in
the inset. The arrows indicate the transitions that are clearly seen in the experiment. The transition from first to fourth harmonic is
shown by the dashed line.}
\end{figure}

Fig. \ref{sjlc8iv} displays the measured $IV$-curve for zero magnetic flux, or maximum $E_J$, for the 4-SQUID sample together with an IV-curve simulated with
$P(E)$-theory. The locations of the peaks were found to depend only on the magnetic flux, not at all on the gate voltages. To properly identify the energy levels
of the SQUID  we measure IV-curves for different magnetic fields. The peak positions as function of applied flux are shown in Fig \ref{sjlc8p}.

The width of the resonance peaks (about 4 $\mu$eV) is smaller than $k_BT = 7 \mu$eV. The width is therefore, either intrinsic or given by
external noise. Our peak widths are thus comparable to or even smaller than what has been observed in similar spectroscopic studies \cite{bibow}.

The peak structure can be qualitatively explained with a 3-resonator model, where one resonator represents all the SQUIDs and the two other come from
the rest of the measurement circuitry; bonding wires and pads. The parameters of the simulation and simulation circuitry are found in
Fig. \ref{sjlc8iv}. The parameters for the SQUID were taken from independent measurements but the parameters of the two other resonator
circuits were fitted to IV-curve. The resistances used in the simulation represent the broadening of peaks due to dissipation and noise.

The $P(E)$-function in Eq.~(\ref{IV}) was calculated using the integral equation approach presented in Ref.~\onlinecite{ingold2}.
The comparison of the IV-curve with the simulation clearly shows that multiphoton excitations are present in the experiment.
However, the energy levels are not equally spaced as would be the case for a classical inductance.

In order to find a better quantitative agreement with the level spacing, the Schr\"odinger equation
(\ref{SCH}) was numerically solved under the assumption that the particle is localized in one of the wells.
The experimental peaks together with the calculated transitions are shown in Fig \ref{sjlc8p}.
The form of the cosine potential decreases the level spacing from the harmonic case. This deviation from the harmonic oscillator
case is largest for the third transition as can be seen in Fig.~\ref{sjlc8p}, where also the transition from the first to fourth
harmonic level is shown for comparison.

The parameter for $E_J$ used in the calculation is about twice the value given by the A-B relation (see Table 1). Typically, $E_J$ is expected to be renormalized
downwards due to the low impedance environment, but in our case it is renormalized upwards.
Similar disagreements between the A-B value and have been reported before \cite{Haviland2,bibow}.
Our model does not, however, explain the double peak structure (see Fig.~\ref{sjlc8p}) found in all the IV-curves. This peak splitting is fairly
constant over the whole measurement range but the position of the double peak is different for the three main transitions and, as can be seen in
Fig.~\ref{sjlc8iv}, the multi-photon excitations given by $P(E)$-theory are not (except for transition (1,1,0)) consistent with the observed peaks.

The 2-point measurements with both one and two SQUID(s) showed similar behavior as the 4 SQUID measurements. The number of SQUIDs in the sample
did not seem to have any significant effect on the IV-curve.  Rather, there are notable differences between the 2-lead and 4-lead samples.
In the 1-SQUID sample with only two leads the current dropped very fast when tuning down $E_J$ (from 360 pA at $\Phi/\Phi_0 = 0$ to 40 pA at
$\Phi/\Phi_0 = 0.4$). This behavior can be explained when considering
that the current through the circuit is given by two rates: the excitation of oscillator modes in the SQUID and their subsequent relaxation, which
depends on the environment seen by the SQUID. In the 2-lead circuits the current is limited by the down relaxation and the effect
can be approximatively explained with the formula \cite{martinis}
\begin{equation}
\Gamma_{\downarrow} = 2 \sum_{l < n}\rm{Re}\{Y(E_{ln}/\hslash)\}(E_{ln}/\hslash) R_Q  |\left\langle l \right\vert
\varphi \left\vert n \right\rangle|^2,
\end{equation}
where the admittance is given by $Y(\omega) = [1/(i\omega C_{Det}) + R_0]^{-1}$ and $C_{Det}$ is the capacitance of the detector junction (0.8 fF) in series
with the resistance of the environment $R_0$ (100 $\Omega$).

\begin{figure}[b]
\includegraphics[width=0.8\linewidth]{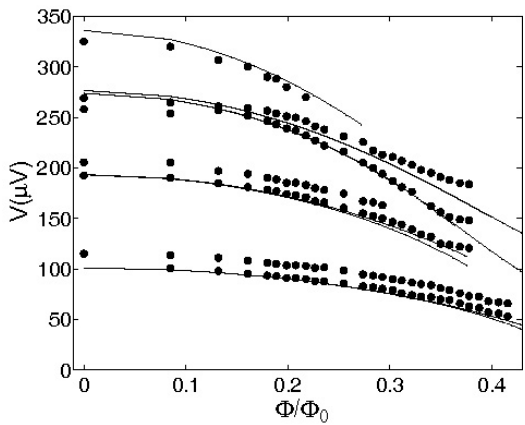}
\caption{\label{sjsqtr7b}Position of the measured resonances for the 1-SQUID sample and the theoretical transitions between band edges
when the wavefunctions are 4$\pi$-periodic. The resonances are grouped into 4 groups, which are transitions between bands 1$\rightarrow$2,
1$\rightarrow$3, 1$\rightarrow$ 4 and 1$\rightarrow$5.  As in the 4-SQUID sample, there is a double peak structure which it not explained by the model.}
\end{figure}

The position of the clearest flux-dependent peaks for the 1-SQUID sample are shown in Fig.~\ref{sjsqtr7b}. In this case, better agreement with the
measured resonances is found when considering the full periodicity of the cosine-potential in Eq.~\ref{SCH}. However, the value
of $E_J$ was taken to be twice the A-B value, as in the case with four SQUIDs. This apparent enhancement of $E_J$ is probably
due to the charging energy, as discussed in Ref.~\onlinecite{zaikin}. The effect according to the theory is, however,
smaller than what we observe.

Because the transitions are due to transfer of Cooper-pairs in the detector junction, the allowed first-order transitions can be found by calculating
the matrix element $|\left\langle l \right\vert e^{-i\varphi} \left\vert n \right\rangle|$, between bands $l$ and $n$. In addition, due to van-Hove
like singularities, the observed transitions are between band edges. The transition between the first and fourth band is clearly
visible as two distinct peaks. The lowest bands are so narrow that they only show up in the width of the resonance peaks. As the $E_J/E_C$ ratio is tuned down,
the life-time of the states grows and this should lead to narrower peaks. But, instead we observe a broadening of the resonances, indicating a broadening of
the bands as expected from theory.

The theory for Josephson junctions \cite{Schon} tells that in order for band formation we need to suppress the ohmic or quasiparticle dissipation
in the environment, which causes the phase to localize. In our system, this suppression is provided by the large quasiparticle resistance of the detector junction.
Therefore, the wavefunctions are 4$\pi$-periodic and each band from the 2$\pi$-periodic case is split into two. The observed transitions are, however, the same
as what would be expected for 2$\pi$-period bands. Consequently, the true periodicity of the bands cannot be resolved in the experiment.

In summary, we have experimentally studied the quantum mechanical energy levels of the Josephson junction. Our results for samples with $E_J/E_C \gg 1$
can qualitatively be described by $P(E)$-theory, consistent with multi-photon excitations in the experiment. The non-linearity of the SQUID systems,
prominent of $E_J \sim E_C$, can be taken into account by considering the exact form of the cosine-potential. Evidence of the existence of
Bloch bands is observed in our 2-lead samples both in the form of van-Hoven like singularities between band edges and a broadening of resonance peaks.
Our results show that a small superconducting junction can be employed as a detector for mesoscopic quantum circuits.

Fruitful discussion with D. Haviland, F. Hekking, F. Wilhelm, G. Sch\"on, J. Siewert, E. Thuneberg,
A. Zaikin and T. Heikkil\"a are gratefully acknowledged. This work was supported by the Academy of
Finland and by the Large Scale Installation Program ULTI-3 of the European Union.

\end{document}